\PassOptionsToPackage{unicode}{hyperref}
\PassOptionsToPackage{hyphens}{url}
\documentclass[
  astrosymb, twocolumn, twocolappendix, tighten]{aastex701}
\usepackage{xcolor}
\usepackage{amsmath,amssymb,amsthm}

\usepackage{iftex}
\ifPDFTeX
  \usepackage[T1]{fontenc}
  \usepackage[utf8]{inputenc}
  \usepackage{textcomp} 
\else 
  \usepackage{unicode-math} 
  \defaultfontfeatures{Scale=MatchLowercase}
  \defaultfontfeatures[\rmfamily]{Ligatures=TeX,Scale=1}
\fi
\usepackage{lmodern}
\usepackage[]{natbib}
\let\oldbibstyle\bibliographystyle
\renewcommand
\makeatletter
\@ifundefined{KOMAClassName}{
  \IfFileExists{parskip.sty}{%
    \usepackage{parskip}
  }{
    \setlength{\parindent}{0pt}
    \setlength{\parskip}{6pt plus 2pt minus 1pt}}
}{
  \KOMAoptions{parskip=half}}
\makeatother
\usepackage{graphicx}
\makeatletter
\newsavebox\pandoc@box
\newcommand*\pandocbounded[1]{
  \sbox\pandoc@box{#1}%
  \Gscale@div\@tempa{\textheight}{\dimexpr\ht\pandoc@box+\dp\pandoc@box\relax}%
  \Gscale@div\@tempb{\linewidth}{\wd\pandoc@box}%
  \ifdim\@tempb\p@<\@tempa\p@\let\@tempa\@tempb\fi
  \ifdim\@tempa\p@<\p@\scalebox{\@tempa}{\usebox\pandoc@box}%
  \else\usebox{\pandoc@box}%
  \fi%
}
\def\fps@figure{htbp}
\makeatother
\providecommand{\tightlist}{%
  \setlength{\itemsep}{0pt}\setlength{\parskip}{0pt}}
\usepackage[]{natbib}
\usepackage{mathptmx,txfonts,tikz,bm}
\usepackage[capitalize]{cleveref}

\newcommand{\annotate}[2]{\begin{tikzpicture}
    \node[anchor=south west,inner sep=0,align=center] (image) at (0,0) {
    #1
    };
    \begin{scope}[x={(image.south east)},y={(image.north west)}]
    #2
    \end{scope}
\end{tikzpicture}}
\def\bibitem{\vskip\bibskip\footnotesize\savebibitem}

\newcommand{\edit}[2]{{\ifnum#1 <  1 %
#2%
\else%
\textbf{#2}%
\fi}}

\usepackage{CJKutf8}
\usepackage{newunicodechar}
\usepackage{xspace}

\newcommand{\mesa}{{\textsc{mesa}}}

\newcommand\gyre{{\textsc{gyre}}}

\defcitealias{ong_semianalytic_2020}{OB20}

\defcitealias{ong_rotation_2022}{OBB22}

\defcitealias{ong_rotation_2023}{OG23}

\newcommand{\chinesename}{{\begin{CJK}{UTF8}{gbsn}(王加冕)\end{CJK}}}

\DeclareRobustCommand{\okina}{%
  \raisebox{\dimexpr\fontcharht\font`A-\height}{%
    \scalebox{0.8}{`}%
  }%
}
\newunicodechar{ʻ}{\okina}

\graphicspath{{./},{./figures/}}

\oldbibstyle{aasjournal}
\usepackage{bookmark}
\IfFileExists{xurl.sty}{\usepackage{xurl}}{} 
\begin{document}

\let\[\relax \let\]\relax 
\DeclareRobustCommand{\[}{\begin{equation}}
\DeclareRobustCommand{\]}{\end{equation}}

\title{Can Asteroseismic Structure Inversions Be Performed in Structure-Dependent Coordinates?}

\correspondingauthor{J. M. Joel Ong}
\email{joel.ong@sydney.edu.au}

\author[orcid=0000-0001-7664-648X]{J. M. Joel Ong \chinesename}
\email{joel.ong@sydney.edu.au}
\altaffiliation{NASA Hubble Fellow}
\affiliation{Institute for Astronomy, University of Hawaiʻi, 2680 Woodlawn Drive, Honolulu, HI 96822, USA}
\affiliation{Sydney Institute for Astronomy (SIfA), School of Physics, University of Sydney, NSW 2006, Australia}

\author[orcid=0000-0002-6163-3472]{Sarbani Basu}
\email{sarbani.basu@yale.edu}
\affiliation{Department of Astronomy, Yale University, P.O. Box 208101, New Haven, CT 06520, USA}

\author[orcid=0000-0003-3953-9532]{Willem Hoogendam}
\email{willemh@hawaii.edu}
\altaffiliation{NSF Graduate Research Fellow}
\affiliation{Institute for Astronomy, University of Hawaiʻi, 2680 Woodlawn Drive, Honolulu, HI 96822, USA}

\author[orcid=0000-0003-4923-6199]{Vincent Vanlaer}
\email{vincent.vanlaer@kuleuven.be}
\altaffiliation{FWO PhD Fellow}
\affiliation{Institute of Astronomy, KU Leuven, Celestijnenlaan 200D, 3001, Leuven, Belgium}

\shortauthors{Ong, Basu, Hoogendam, and Vanlaer}
\shorttitle{Generalised Inversions}
\begin{abstract}
Yes. Unlike other applications of observational asteroseismology, ``structure inversions'' constrain the physical properties of stellar interiors in a model-independent fashion. However, existing techniques --- which parameterise these quantities as functions of the physical radial or mass coordinate --- break down when applied to stars which differ substantially from the Sun. These difficulties may be overcome by operating in coordinate systems that have long been known to more naturally suit the physical characteristics of the measured normal modes. We derive expressions for transforming inversion kernels in the acoustic and buoyancy radial coordinates, rather than in the physical radius, and make available a numerically performant implementation. These modified inversions directly address several specific known shortcomings of existing inversion procedures. Using the buoyancy radius in gravity-mode and mixed-mode pulsators permits meaningful comparisons of stars and models with differently-sized convective and radiative zones, which defeat standard inversions. Even in pressure-mode oscillators, inversions in the acoustic radial coordinate eliminate the methodological requirement for the mass and radius of the true star being needed to be known in advance.
\end{abstract}

\def\sectionautorefname{\S}
\def\subsectionautorefname{\S}
\def\subsubsectionautorefname{\S}

\section{Introduction and Motivation}\label{introduction-and-motivation}

The most common applications of observational asteroseismology are built on forward modelling, in which one minimises discrepancies between the observed normal-mode frequencies of a pulsating star, and those of some best-fitting model of its structure. However, more advanced ``inversion'' techniques permit the mode frequencies of a star to directly constrain the differences between its interior structure and that of such a model. The measurements of e.g.~rotation rate, density, or sound speed that are returned from such inversion procedures are essentially independent of the forward model, as long as the structure of the forward model is close enough to that of the true star.

For this to work, one must assume that the frequency differences between the true star and its reference model take the form
\[
{\delta \omega_i \over \omega_i} \sim \sum_j \int_0^1 K_{i, j}(x)\delta u_j(x) \ \mathrm d x, \label{eq:kernel}
\]
where the integration runs over the fractional physical radial coordinate, and the sum runs over all physical quantities that induce changes to the mode frequencies independently of each other. The kernel function \(K_{i,j}(x)\) specifies how sensitive the \(i\)\textsuperscript{th} normal-mode frequency is to a perturbation in the \(j\)\textsuperscript{th} physical quantity localised at position \(x\). Through it, \cref{eq:kernel} specifies a linear map between the frequency differences to perturbations \(\delta u(x)\) to the physical quantities under consideration. The objective of an inversion is to solve for the latter as linear combinations of the former: in this fashion, \cref{eq:kernel} constitutes a linear inverse problem.

\cref{eq:kernel} is written so that the unknown quantities are Eulerian structure differences, taken at the same fractional radius in the model and the true star. Studying Lagrangian differences, taken instead at the same mass coordinate, has been known be instructive for studying the Sun in particular \citep{jcd_solar_1997}, where the total mass is known. Inversions for stars other than the Sun pose a further issue: unlike the Sun, the mass and radius of a star is not generally known independently of asteroseismology, and so the stellar model used as a reference for inversions may not necessarily have the same mass or radius as the true star. In order to compare frequencies and physical quantities between stellar structures with different masses and radii, the frequency differences on the left-hand-side of \cref{eq:kernel}, or those of physical quantities on the right-hand-side, must be evaluated with respect to dimensionless quantities: typically, as scaled with respect to the dynamical frequency \(\omega_0 = \sqrt{GM/R^3}\), and stellar radius \(R\) as characteristic time- and lengthscales. Asteroseismology itself is often the primary means by which these quantities are determined, but such estimates are \emph{properties of the forward models}, and not those of the true star as needed for this scaling. Without independent measurements of the stellar mass and radius, even the inputs to the inversion problem are difficult to construct in the first place. As such, structure inversions have only successfully been performed on cool solar-like oscillators, similar to the Sun \citep[e.g.][]{bellinger_radiative_2017, bellinger_convective_2019, buldgen_16cyg_2022, buchele_radiative_2024, buchele_convective_2025}.

In lieu of either the fractional physical radius or mass, other coordinate systems (and accordingly choices of natural units) have historically proven more suitable or convenient for analysing the physical properties of normal modes \citep[e.g.~for asymptotic analysis:][]{tassoul_asymptotic_1980, tassoul_second_1990, roxburgh_asymptotic_1994}. For example, the roots of pressure-mode (p-mode) eigenfunctions --- and thus the sensitivity of kernels generated from them --- are heavily concentrated near the surface of the star. It has long been known that kernels constructed in the fractional acoustic radial coordinate \(t(r)/t(R)\), where
\[t(r) = \int_0^r{\mathrm d r' \over c_s},\label{eq:t}\]
would have the desirable property of uniformly distributing the sensitivity of such kernels throughout the unit interval \citep{thompson_seismic_1993, pijpers_sola_1994}. Here \(c_s\) is the adiabatic sound speed.
Such uniform sensitivity would also greatly simplify the construction of resolution elements or target kernels for structure inversions.

Similarly, asymptotic analysis of gravity modes (g-modes), in stars hosting radiative zones, reveals that the fractional buoyancy radial coordinate \(b(r)/b(R)\), where
\[b(r) = \int_0^r{N \over r'}\ \mathrm d r',\label{eq:b}\]
is a natural coordinate in which the sensitivity of the normal-mode eigenfunctions is most uniformly distributed over the unit interval; this property has been used in recent attempts at asymptotic inversions \citep{briganti_predictions_2025, guo_glitch_2025, guo_inferring_2025}. Here \(N\) is the Brunt-Väisälä frequency. This suggests the use of this coordinate for g-mode inversions, avoiding a known issue where even stars with identical masses and radii to a reference model might possess differently-sized convective and radiative zones from it, in the fractional physical radial coordinate. Under the standard choice of scaling, this would otherwise cause the kernels of the reference model to probe significantly different regions than where g-modes in the true star may propagate \citep{vanlaer_feasibility_2023}.

Unfortunately, inversions in these physically motivated radial coordinates, using fully nonasymptotic kernels, have so far only been possible to employ for studying rotation \citep[e.g.~as in][]{ong_zvrk_2024}, and not structure. As we shall see, this is because the structure inversion kernels transform nontrivially between coordinate systems if these choices of coordinates themselves depend on the quantities being perturbed. As a result, the potential for structure inversions to be carried out using them has never previously been examined. This capability would, however, be generally useful for carrying out asteroseismic structure inversions, and particularly so for stars that are not Sun-like --- where techniques designed for the Sun have been shown to work the least well.

Therefore, in this work, we derive expressions for these transformations --- both generally in \autoref{sec:lagrangian}, and for specifically p-mode oscillators in \autoref{sec:acoustic} and g-mode ones in \autoref{sec:buoyancy}. We further examine how they may modify existing asteroseismic inversions (\autoref{sec:inversions}).

\section{Structure-Dependent Coordinates under Structural Variations}\label{structure-dependent-coordinates-under-structural-variations}

\label{sec:lagrangian}

\newcommand{\LL}{{\ensuremath\mathcal{L}}}
\newcommand{\DD}{{\ensuremath\mathcal{D}}}

We will formulate inversion techniques based on a description of the normal-mode frequencies \(\omega\) of a pulsating star as satisfying an eigenvalue equation, which \citep[following][]{ong_rotation_2022} we shall express in matrix form as
\[
\left(\omega^2 \mathbf{D} + \mathbf{L}\right)\mathbf{c} = 0.\label{eq:eig}
\]
These matrices collect matrix elements of the identity operator and the wave operator, respectively, as evaluated with respect to a chosen set of Lagrangian displacement basis functions. These matrix elements are evaluated by taking the natural inner product, in the sense of integrating against the mass coordinate. E.g.,
\[
D_{ij} = \left<\vec\xi_i, \hat{H}_\mathbf{D} \vec\xi_j\right> = \int \vec\xi_i^* \cdot (\hat{H}_\mathbf{D} \vec\xi_j)\ \mathrm d m,
\]
where \(\hat{H}_D\) is the differential operator whose matrix elements, evaluated in this fashion, are collected in \(\mathbf{D}\). We have adopted the convention that the basis functions are each normalised as \(\left<\xi_i, \xi_i\right> = 1\). In particular, where the basis functions \(\xi_i\) are the natural basis of normal modes, then \(D_{ij} = \delta_{ij}\); however, this is not generally true. Each eigenvector \(\mathbf{c}\) of \cref{eq:eig} specifies a linear combination of these basis functions which yields that normal-mode eigenfunction.

Suppose the interior of the star (and therefore the normal-mode basis functions \(\xi_k\), frequencies \(\omega_k\), and operators \(\hat{H}_\mathbf{M}\)), are perturbed such that \(\hat{H}_\mathbf{M} \mapsto \hat{H}_\mathbf{M} + \lambda\hat{V}_\mathbf{M}\) for each matrix \(\mathbf{M}\). Rayleigh-Schrödinger perturbation analysis dictates that, when the matrix elements are expressed in the natural basis of normal modes, the mode frequencies are altered as
\[
\omega^2_{0,k} \mapsto \omega^2_k = \omega^2_{0,k} - \lambda(\delta L_{kk} + \omega^2_{0,k} \delta D_{kk}) + \mathcal{O}(\lambda^2).\label{eq:rsperturb}
\]
\cref{eq:kernel} follows from truncating \cref{eq:rsperturb} to leading order in the perturbation, with the kernel function being the integrands required to evaluate various on-diagonal matrix elements. Off-diagonal matrix elements become relevant if the perturbation should be large, so that this representation of the matrices appearing in \cref{eq:eig} ceases to be diagonally dominant --- but those appear only at higher order in \(\lambda\), at least when expressed in the natural basis of normal modes. While, in this work, we focus only on relating the diagonal matrix elements to structural perturbations, the considerations we present are generally applicable to the off-diagonal terms as well.

\subsection{Recap of Variational Calculus}\label{recap-of-variational-calculus}

Normal modes in asteroseismology are often formulated as emerging from a variational principle, in the sense of their wave equations being the Euler-Lagrange equation associated with some ``action'' functional \citep[e.g.][]{chandrasekhar_variational_1964, lyndenbell_stability_1967}. Solutions to the Euler-Lagrange equation are functions that \emph{minimise} such an action functional. However, rather than searching for minima, we may also use variational calculus to keep track of how physical perturbations, treated as functions of position, change the value of a functional away from such local minima. Since this application is not often employed in the asteroseismic literature, we first recount some basic definitions.

Consider a functional \(S[f_k(x), x]\) that depends on a coordinate \(x \in [0, 1]\), some field quantities \(f_k(x)\), and other quantities derived from the fields, e.g.~including their derivatives \(f_k'(x)\) and antiderivatives \(I_{f_k}(x) = \int_0^x  \mu_k(x') f_k(x')\ \mathrm d x'\) with respect to the coordinate. We may express \(S\) in the form of an integral against a functional density \(\LL\) as
\[S[f_k(x)] = \int_0^1 \mathcal{L}[x, f_k(x), \ldots]\ \mathrm d x.\]
Consider the simplified case of there being only a single field \(f\). Under a perturbation \(f(x) \mapsto f(x) + \lambda \delta f(x)\) (and likewise for its derivatives and antiderivatives), we may compute the variation to the value of the functional \(S\) itself as
\[\begin{aligned}S[f + \lambda \delta f] &- S[f]  \equiv \delta S[f] \\&= \lambda\int_0^1 \left\{{\partial \LL \over \partial f}\delta f + {\partial \LL \over \partial f'}\delta f' + {\partial \LL \over \partial I_f}\delta I_f\right\}\ \mathrm d x + \mathcal{O}\left(\lambda^2\right).\end{aligned}\]
Integrating by parts, we obtain
\[
\begin{aligned}
\delta S[f] &= \lambda\left[\left({\partial \LL \over \partial f'}\delta f(x)\right)_0^1 + \left(\delta I_f(x)\int_0^x{\partial \LL \over \partial I_f(x')}\ \mathrm d x' \right)_0^1\right.\\  &\left. + \int_0^1 \left\{{\partial \LL \over \partial f} - {\mathrm d \over \mathrm d x} {\partial \LL \over \partial f'} - \mu(x)\int_0^x {\partial \LL \over \partial I_f(x')}\ \mathrm d x'\right\} \delta f(x)\ \mathrm d x\right]  + \mathcal{O}\left(\lambda^2\right)
\\ & \equiv \lambda \int_0^1 {\delta S \over \delta f} \delta f(x)\ \mathrm d x  + \mathcal{O}\left(\lambda^2\right),
\end{aligned}\label{eq:funcder}
\]
where \({\delta S \over \delta f}\) is, in compact notation, the \emph{functional derivative} (or \emph{variational derivative}) of \(S\) with respect to the function \(f(x)\). The boundary terms are usually ignored by formulating the problem in such a way that \(\delta f(x)\) and \(\delta I_f(x)\) vanish at the endpoints. For example, when \(I_f\) is the mass coordinate and \(f\) is the density, we may formulate the problem so that \(\delta I_f(1) = 0\) to conserve the total mass.

Formally, each matrix element of \cref{eq:eig} may be treated as such a functional, mapping functions of the radial coordinate to single numbers. While the wave equation may be recovered by varying them with respect to the eigenfunctions, \cref{eq:kernel} is instead obtained by varying the on-diagonal matrix elements with respect to structural quantities, such as the density and sound speed. The sensitivity kernels there can then be seen simply to be the functional derivatives of these matrix elements, with respect to these structural quantities.

\subsection{Changes of Coordinates}\label{sec:changes}

Of particular interest to us is the possibility of changing coordinates, say from \(x\) to some other variable \(y \in [0, 1]\). While this has been done in the specific case of the mass coordinate, in \citet{jcd_solar_1997}, we seek a general expression applicable to other coordinate systems. Let's say the two coordinates are related by some smooth, monotonic function so that \(y = y(x) \iff x = x(y)\). If \(x=0\) where \(y = 0\), we might construct \(y(x)\) to be of the form \(h(x) / H\), where \(h = \int_0^x J(x)\ \mathrm d x'\) and \(H = h(1)\), for some weight function \(J(x) > 0\). Alternatively, we may choose \(H\) to be some constant whose value is chosen to coincide with the value of \(h(1)\), but not otherwise assumed to change when \(y\) is varied. We will refer to the first case as where we would maintain a \emph{fixed} outer boundary in the \(y\) coordinate upon variation, vs.~if we were to leave the outer boundary \emph{free} to change under the variation.

In principle, we would take functional derivatives with respect to new functions \(\tilde{f}(y)\) defined so that \(\tilde{f}(y(x)) = f(x)\), and compute
\[
\delta S = \int {\delta S\over \delta \tilde f}\delta\tilde f(y)\ \mathrm d y
\]
by replacing \(\int\LL \mathrm d x\) with \(\int \LL{\mathrm d x \over \mathrm d y} \mathrm d y \equiv \int \tilde{\LL} \mathrm d y\) in \cref{eq:funcder}. In practice, doing so \emph{ab initio} is typically more cumbersome than finding it from \({\delta S \over \delta f(x)}\), which often is already known.

It might be tempting to identify \({\delta S \over \delta f(x)}\) with \({\delta S \over \delta \tilde f(y)}{\mathrm d y \over \mathrm d x}\) --- i.e.~to assume that the two are related only by rescaling to account for the difference in integral measure incurred by changing coordinates. This does indeed yield the same variation \(\delta S\) when \(\delta \tilde{f}(y(x)) = \delta f(x)\). However, when the coordinate transformation depends on the field \(f\), then \(\delta \tilde{f}(y(x)) \ne \delta f(x)\) even if \(\tilde{f}(y(x)) = f(x)\), since the coordinate \(y(x)\) itself depends on the variation \(\delta f\). Instead, since \(\tilde{f}(y(x)) = f(x)\) must remain valid even upon variation, we have by the chain rule that
\[
\begin{aligned}
f(x) + \lambda\delta f(x) &= (\tilde{f} + \lambda\delta\tilde f)([y + \lambda \delta y](x)) \\
\implies\lambda \delta \tilde f(y(x)) &= \lambda \left[\delta f(x) - {\mathrm d \tilde f\over \mathrm d y} \int {\delta y(x) \over \delta f(x')} \delta f(x')\ \mathrm d x'\right] + \mathcal{O}\left(\lambda^2\right).
\end{aligned}\label{eq:lagrange-to-euler-single}
\]
In principle, one might also expect that the Jacobian of the coordinate transformation should also be varied; however, this cancels out with the variation to the value of the integration measure \(\mathrm d y\) itself.

Here \(y(x)\) is itself treated as a functional that depends on \(f\), and its functional derivative results in an additional term. This term appears for similar reasons to that which appears when one relates Lagrangian fluid perturbations to Eulerian ones. Perturbations parameterised by \(x\) are ``Eulerian'' in the sense of depending on a coordinate system that is not varied by \(\delta f\), whereas those parameterised with respect to \(y\) are ``Lagrangian'', as their integration coordinate is field-dependent (and thus ``advected'' under variations to the structural fields). We will refer to the former as Eulerian perturbations, and the corresponding kernels of \cref{eq:kernel} as Eulerian kernels. While it is tempting to refer to the latter as ``Lagrangian'' kernels, this is already a term of art in the asteroseismic literature \citep[referring to kernels in the mass coordinate, a la][]{jcd_solar_1997}; we will instead refer to \(\tilde{K}\) as being ``modified'' kernels in a given coordinate.

\subsection{Inversion Kernels under Changes of Coordinates}\label{inversion-kernels-under-changes-of-coordinates}

Although \cref{eq:lagrange-to-euler-single} specifies \(\delta\tilde{f}(y)\) in terms of \(\delta f(x)\), our objective is to construct inversion kernels integrated against \(y\) out of those integrated against \(x\) --- that is, to find \(\delta S \over \delta \tilde f(y)\) in terms of \(\delta S \over \delta f(x)\). In order to do this, we require an inverse relation, expressing \(\delta f(x)\) in terms of \(\delta\tilde{f}(y)\). Unfortunately, \cref{eq:lagrange-to-euler-single} is a Volterra integral equation of the first kind, for which in general no closed-form analytic solutions exist.

In the case of multiple fields --- i.e.~where more than one quantity may be perturbed simulaneously, as in \cref{eq:kernel} --- \cref{eq:lagrange-to-euler-single} generalises naturally\footnote{Although we have not adopted any nonstandard notation in \cref{eq:lagrange-to-euler}, it is unfortunate that this equation employs the letter \(\delta\) to mean four different things: \(\delta f_k\) being a variation to \(f_k\), \(\delta_{kl}\) being the Kronecker delta symbol, \(\delta(x-x')\) being the Dirac delta distribution, and \(\delta S \over \delta f\) being the variational derivative of the functional \(S\) with respect to the function \(f\).} to
\[
\delta \tilde f_k(y(x)) = \sum_l \int \mathrm d x' \left[\delta_{kl}\ \delta(x - x') - {\mathrm d \tilde f_k\over \mathrm d y} {\delta y(x) \over \delta f_l(x')} \right]\delta f_l(x').
\label{eq:lagrange-to-euler}
\]
Let us suppose that there exist integral kernels \(\tilde{K}_i\) for each \(\tilde{f}_i\), which we also demand satisfy \cref{eq:kernel}. By \cref{eq:lagrange-to-euler}, these are related to the ordinary kernels \(K_i\) through an integral equation of the form
\[
\begin{aligned}
\delta S &= \sum_k \int \tilde K_k(y) \delta \tilde f_k(y)\ \mathrm d y \\
&= \sum_{kl} \iint \tilde K_k(y) \left[\delta_{kl}\ \delta(x - x') - {\mathrm d \tilde f_k\over \mathrm d y} {\delta y(x) \over \delta f_l(x')} \right]\delta f_l(x')\  \mathrm d x'\mathrm d y \\
&= \sum_k \int K_k(x) \delta f_k(x) \mathrm d x.
\end{aligned} 
\]
We can see that these \(\tilde K_i\) satisfy a conjugate set of Volterra integral equations to the perturbations \(\tilde \delta f_i\). Explicitly,
\[
K_k(x') = \sum_{l} \int {\mathrm d y \over \mathrm d x} \tilde K_l(y(x)) \left[\delta_{kl}\ \delta(x - x') - {\mathrm d \tilde f_l\over \mathrm d y} {\delta y(x) \over \delta f_k(x')} \right]\  \mathrm d x \\
\]
Except in special cases, solutions can only be found approximately. For example, we might apply the technique of iterative approximations by Neumann series, where we use the original kernels as an initial guess \(\tilde{K}_k^{(0)}(y(x)) = {\mathrm d x \over \mathrm d y} K_k(x)\), and construct successive iterations to \(\tilde{K}_k\) as
\[
{\mathrm d y \over \mathrm d x}\tilde{K}_k^{(i)}(y(x)) = K_k(x) + \sum_l \int {\mathrm d \tilde f_l \over \mathrm d y}(x') {\delta y(x') \over \delta f_k(x)}\left({\mathrm d y \over \mathrm d x}\tilde{K}^{(i-1)}_l(y(x'))\right)\ \mathrm d x'.\label{eq:iteration}
\]
Iterative procedures like \cref{eq:iteration} are not guaranteed to converge. However, we find that these series solutions do converge quite rapidly for the coordinate transformations that we consider in this work. We examine this in more detail (accounting also for numerical discretisation) in \autoref{sec:convergence}.

\section{Physical Coordinate Systems}\label{physical-coordinate-systems}

To illustrate this procedure, we now examine how commonly-used inversion kernels, of the kind appearing in \cref{eq:kernel}, transform under various choices of structure-dependent coordinate systems. In particular, as described in our introduction, we will examine p-mode kernels constructed in the acoustic radial coordinate, and g-mode kernels in the buoyancy radial coordinate.

\subsection{The Acoustic Radial Coordinate}\label{sec:acoustic}

The acoustic radius, \cref{eq:t}, is the natural radial coordinate for p-modes. Most of the time, only two thermodynamical variables are required to completely describe the perturbation to the stellar structure \citep[e.g.][]{dappen_new_1991, gough_inversion_1991, kosovichev_inversion_1999}. For p-mode oscillators, the density \(\rho\) and sound speed \(c_s\) are a common choice of kernel pair, so that
\[
\delta S = {\delta\omega\over \omega} = \int_0^1 K_\rho\ {\delta\log \rho}\ \mathrm d x + \int_0^1 K_{c_s}\ {\delta\log c_s}\ \mathrm d x.\label{eq:rhoc}
\]
In practical applications, \cref{eq:rhoc} is not used directly, as both the frequency differences and these structural variables must also be subjected to nondimensionalisation, which we explain in more detail in \autoref{sec:inversions}. While \citet{pijpers_sola_1994} propose the use of the acoustic radial coordinate for inversions, they were not able to derive the transformation from \(K\) to \(\tilde{K}\) that we have described, and so the kernels they discuss are simply the Eulerian kernels \(K\), rescaled by the integration measure of the coordinate transformation.

We illustrate, in the left column of panels of \cref{fig:rescaled}, both these standard Eulerian kernels \(K\) (with the blue dotted curves), and the transformed ones \(\tilde{K}\), under both a fixed outer boundary (orange dash-dotted curve) and free outer boundary (solid gray curve) in the transformation, for kernels in both of these quantities. These were calculated for the kernels of the \(\ell = 1, n_p = 16\) mode of Model S \citep{jcd_current_1996}. The partial functional derivative of \(y\) with respect to the density vanishes, and this is reflected in the fact that, for density perturbations, the Eulerian and modified kernels coincide. Instead only the \(c_s\) kernel is substantially modified. Counterintuitively, the outer boundary condition of the coordinate transformation primarily determines the shape of the transformed kernels close to the centre of the star at \(t = 0\), rather than at the outer boundary. While the Eulerian sound-speed kernel is strictly positive, a free boundary causes the modified kernel to oscillate around zero, and a fixed boundary causes a systematic positive offset close to the centre of the star. As we will see, this causes significant problems when attempting to construct localisation kernels; for practical applications, we will focus primarily on a free outer boundary.

\begingroup
\renewenvironment{figure}{\begin{figure*}}{\end{figure*}}

\begin{figure}
\centering
\includegraphics[width=0.49\linewidth,height=\textheight,keepaspectratio]{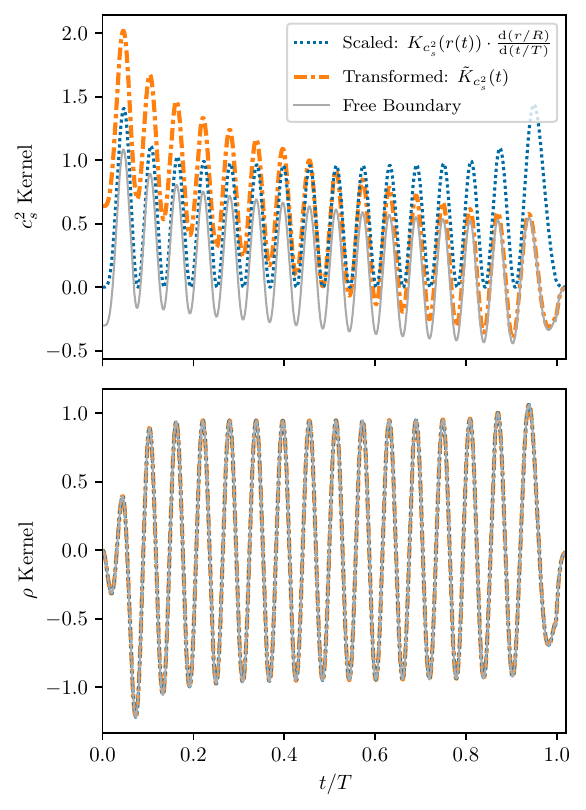}
\includegraphics[width=0.49\linewidth,height=\textheight,keepaspectratio]{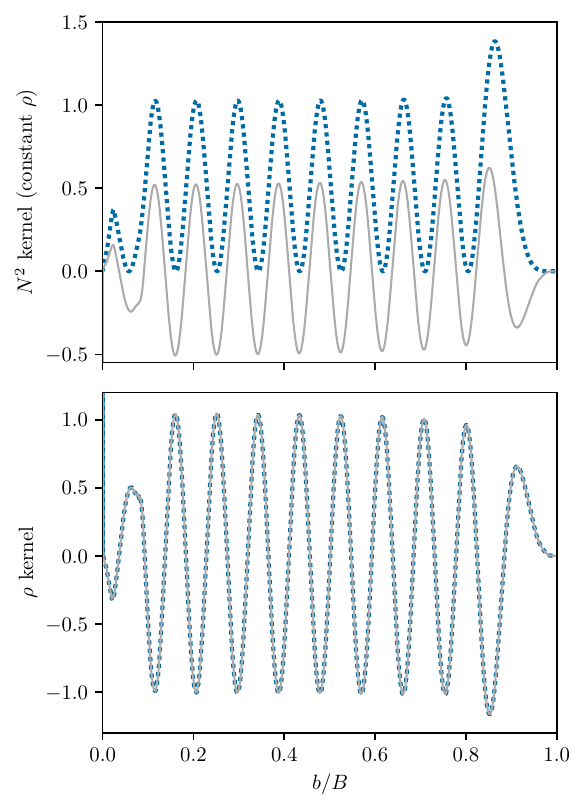}
\caption{Illustration of differences between \(\tilde{K}(y)\) (orange dash-dotted and gray solid curves) and \(K(x(y)){\mathrm d x \over \mathrm d y}\) (blue dotted curve) for two different stellar models. In the left column of figures \citep[constructed using Model S:][]{jcd_current_1996}, we illustrate how the kernel pair \(K_\rho\) and \(K_{c_s}\) change under the transformation from \(x = r/R\) to \(y = t/T\). For comparison, the Eulerian kernel (only rescaled by the integral measure) is shown with the blue dotted curve. Since the change of coordinates only depends on \(c_s\), the \(\rho\) kernel is unchanged apart from rescaling. However, the \(c_s\) kernel is significantly modified by this transformation. We show both kernels constructed so that the sound-speed perturbation does not alter value of \(y\) at \(x = 1\) (orange dash-dotted curve), and kernels constructed without this constraint (gray solid curve). In the right column of figures \citep[constructed using an SPB model from][]{vanlaer_feasibility_2023}, we show the corresponding transformation applied to g-mode structure kernels in the buoyancy radial coordinate \(y = b/B\), using the Brunt-Väisälä frequency and density as perturbation quantities, using the same colours and line style as for the p-modes. \label{fig:rescaled}}
\end{figure}

\endgroup

\subsection{The Buoyancy Radial Coordinate}\label{sec:buoyancy}

Likewise, the buoyancy radial coordinate, \cref{eq:b}, is the natural radial coordinate for describing g-modes. Motivated by the p-mode case, we choose \((\rho, N^2)\) as our preferred pair of dynamical variables, rather than \((\rho, c_s)\) as in \cref{eq:rhoc}, so that we have again that \(\delta y \over \delta \rho\) vanishes when \(y = b(r) / B\). The calculation of the \((\rho, N^2)\) kernel pair is known to suffer from numerical instabilities; we discuss how we addressed these in order to calculate these kernels in \autoref{sec:N2kernels}. Moreover, in convective regions of a star, the Brunt-Väisälä frequency is (at least formally) 0, which interferes with the formal definition of the buoyancy radial coordinate, \cref{eq:b}, causing it (and the reparameterised kernels) to become singular in each convection zone, at least in principle. Accounting for these singularities would significantly complicate both the implementation of an inversion procedure, and our interpretation of its results.

However, it is possible to recover non-singular behaviour using a modified definition for the buoyancy radial coordinate. As we motivate in more detail in \autoref{sec:deltafuncts}, the Brunt-Väisälä frequency does not actually numerically go to zero in the convection zones of stellar models, but rather takes very small (potentially negative) values. This allows us to define
\[
b_\text{model}(r) = \int_{0}^{r} {\sqrt{\left|N^2_\text{model}\right|}\over r}\mathrm dr,\label{eq:bmod}
\]
where the integral runs over the entire star.
This formulation has the advantage that the integrand is zero only at a finite number of locations (rather than everywhere within a convection zone), which results in a one-to-one (and therefore invertible) map between \(b\) and \(r\). It is also much simpler to implement than the piecewise description of \cref{eq:cz}. We use this definition of \(b\) for all calculations shown for \(\tilde{K}\) in the buoyancy radius. With respect to this modified buoyancy coordinate, for g-mode kernels in particular (for which inversions in the buoyancy coordinate are relevant in the first place), we find the contributions from these singularities not to be significant. We explain this in more detail in \autoref{sec:deltafuncts}.

We illustrate the buoyancy-radius transformation to g-mode kernels in the right column of figures in \cref{fig:rescaled}. These calculations were applied to the the \(\ell = 1, n_g = 10\) mode of the reference model considered in \citet{vanlaer_feasibility_2023}. Again, primarily the \(N^2\) kernel is modified by the coordinate transformation, with the qualitatively similar effect of turning a strictly positive Eulerian sensitivity kernel into a modified one that oscillates around zero.

\section{Applications to Structure Inversions}\label{sec:inversions}

\subsection{p-modes: Solar Structure Inversions}\label{sec:solar}

Existing inversion techniques already apply well to Sun-like stars, so we choose to demonstrate our modified technique on helioseismology. In particular, we consider these modifications applied to model-model inversions using the method of Subtractive Optimal Localised Averages \citep[SOLA:][]{pijpers_sola_1994}.
More details about the SOLA procedure can be found in e.g. \citet{basu_global_2016}. We apply it using Model S \citep{jcd_current_1996} as our reference model, and using the photospheric-mixture standard solar model of \citet{magg_observational_2022} as the target stellar structure, limiting our attention to modes of \(\ell < 400\) and \(1100 < \nu < 3000\ \mu\mathrm{Hz}\), with no modifications to the procedure except for constructing target kernels that are localised at uniformly-spaced steps, and with uniform widths, in the acoustic rather than physical radial coordinate. We focus on differences in these target kernels as constructed using the Eulerian kernels, and both of the modified kernels described in \autoref{sec:acoustic}. Moreover, since only the sound-speed kernels are substantially modified by the coordinate transformation, we will focus on sound-speed inversions.

\begin{figure}
\centering
\pandocbounded{\includegraphics[keepaspectratio,alt={Localisation error of SOLA averaging kernels: the centers of probability masses for localisation kernels constructed out of linear combinations of basis kernels are shown on the vertical axis, and these are shown as a function of the original target location in the SOLA procedure. Differently-coloured curves show different sets of basis kernels used for constructing these localisation kernels. The black curve shows centres of sensitivity for kernels constructed in the physical radial coordinate, with both target and actual locations mapped to the acoustic radial coordinate. The dashed gray line on the diagonal indicates the line of equality. In the background, we show the SOLA averaging kernels associated with the modified kernels with free boundary conditions, with mesh points masked out (shown in white) where the kernels are more than 1/3 of their values at the main peak. Each vertical strip of pixels corresponds to a different localisation kernel, positioned horizontally at its target acoustic radius. Values are shown on a symmetric logarithmic colour map, with positive values shown in green, and negative values in purple. }]{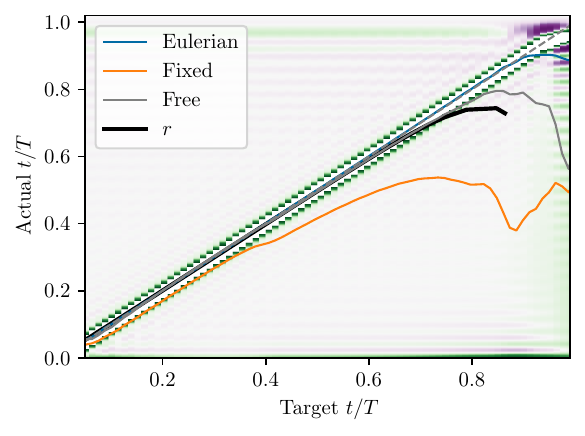}}
\caption{Localisation error of SOLA averaging kernels: the centers of probability masses for localisation kernels constructed out of linear combinations of basis kernels are shown on the vertical axis, and these are shown as a function of the original target location in the SOLA procedure. Differently-coloured curves show different sets of basis kernels used for constructing these localisation kernels. The black curve shows centres of sensitivity for kernels constructed in the physical radial coordinate, with both target and actual locations mapped to the acoustic radial coordinate. The dashed gray line on the diagonal indicates the line of equality. In the background, we show the SOLA averaging kernels associated with the modified kernels with free boundary conditions, with mesh points masked out (shown in white) where the kernels are more than 1/3 of their values at the main peak. Each vertical strip of pixels corresponds to a different localisation kernel, positioned horizontally at its target acoustic radius. Values are shown on a symmetric logarithmic colour map, with positive values shown in green, and negative values in purple. \label{fig:loc}}
\end{figure}

In \cref{fig:loc}, we show the centres of sensitivity (in the sense of a kernel-weighted average location) for SOLA localisation kernels as a function of the target acoustic radius at which they are constructed. Different sets of basis kernels have different localisation properties, even in the same (acoustic) radial coordinate.

\paragraph{Eulerian Kernels}

Interestingly, when parameterised with respect to the acoustic radial coordinate, the Eulerian kernels are capable of forming resolution elements at positions closer to the surface than when applying SOLA using kernels in the physical radial coordinate. As a reference we also show, using the black curve, the centres of sensitivity of localisation kernels constructed using ordinary kernels in the physical radial coordinate, with both target and actual locations mapped to equivalent acoustic radial coordinate. Empirically, we found it easier to construct SOLA localisation kernels out of the Eulerian kernels than the modified ones, particularly close to the surface.

\paragraph{Fixed-boundary Modified Kernels}

For the fixed-boundary kernels, we find that localisation kernels constructed from them suffer from severe ringing artifacts near the surface, irrespective of target radius. We were unable to construct suitable localisation kernels for sound-speed perturbations while simultaneously suppressing sensitivity to density perturbations.

\paragraph{Free-boundary Modified Kernels}

Upon initial inspection, the centres of sensitivity for the free modified kernels appear to diverge systematically from their target locations at large acoustic radii, in much the same fashion as kernels constructed in the physical radial coordinate. To better understand this, we show each free-boundary modified kernel as a vertical column of pixels in the background of \cref{fig:loc}. Inspecting the shapes of these kernels shows that the kernels remain well-localised at the target radius; these offsets are instead mostly driven by the emergence of a single positive sidelobe localised at \(t = 0\) (in addition to some ringing near the surface). While both the Eulerian sound-speed and density kernels vanish at the centre, the modified sound-speed kernels pick up a nonzero bias (as can be seen in the left column of \cref{fig:rescaled}), which is not accounted for in the standard SOLA procedure. On one hand, this means that a modified procedure using these kernels may require a further Lagrange multiplier to be introduced, in order to suppress sensitivity at \(t = 0\) when constructing localisation kernels close to the surface. On the other hand, this also means that this set of kernels may provide sensitivity to structure at \(t = 0\), which is otherwise not the case with the standard set of Eulerian kernels. We discuss this central sensitivity in \autoref{sec:centre}, where we find that this central sensitivity allows the inferred sound-speed profile returned from the inversion procedure to be relatively larger or smaller on average than the reference model, and/or to imply a different total stellar radius.

\begin{figure}
\centering
\annotate{\pandocbounded{\includegraphics[keepaspectratio]{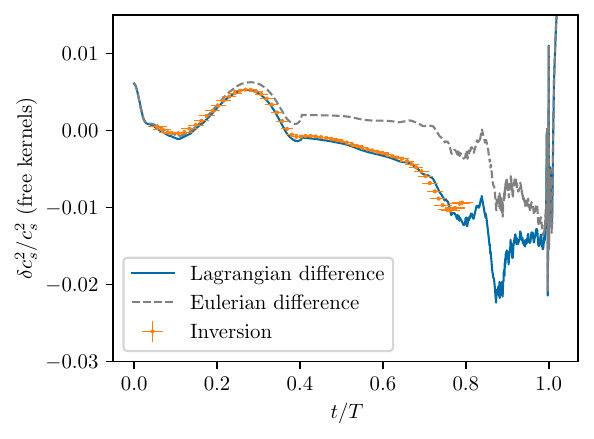}}}{\node at (.23, .9){\textbf{(a)}};}
\annotate{\pandocbounded{\includegraphics[keepaspectratio]{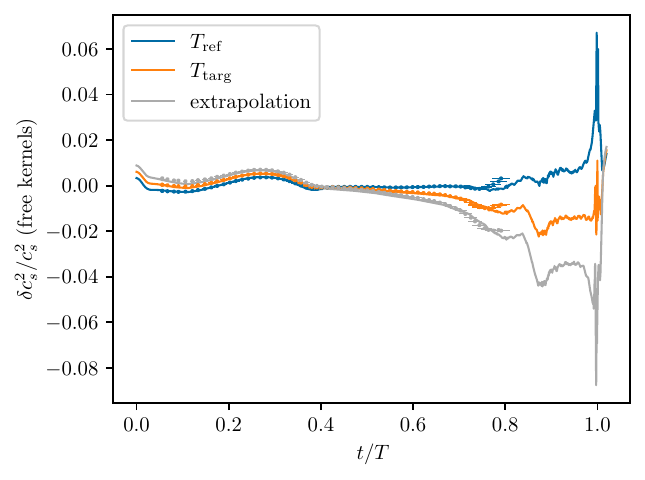}}}{\node at (.23, .23){\textbf{(b)}};}
\caption{Our modified kernels permit model-model inversions in the acoustic radial coordinate for relative differences in the squared sound speed, evaluated at matching \(t/T\) rather than \(r/R\). \textbf{(a)}: SOLA inversion results using basis kernels with free outer boundary conditions. The inverted structure differences trace out the ground-truth differences at matching acoustic radial coordinate, \(\delta c_s(t/T)\) (blue solid curve), rather than the Eulerian differences, \(\delta c_s(r/R)\) (gray dashed curve). \textbf{(b)} The accuracy of these inversion results is insensitive to the choice of characteristic timescale \(\tau\) assigned to the target model. In particular, we compare ground truths (lines) and inversion results (points with error intervals) as derived with \(\tau_\text{targ} = T_\text{ref}\) (blue), \(\tau_\text{targ} = T_\text{targ}\) (orange), and \(\tau_\text{targ} = 2T_\text{targ} - T_\text{ref}\) (i.e.~a linear extrapolation --- gray). While different adopted values change the actual inversion results (data points), they also change the ground truth curves by modifying the acoustic radii in the target model at which differences against the reference model are evaluated.\label{fig:pinversions}}
\end{figure}

To illustrate another advantage of inversions in the acoustic radial coordinate, we examine the results of these inversions using the free modified kernels. Dimensional analysis of the pulsation equations yields that only two characteristic quantities are required to isolate the scale-free behaviour of stellar oscillations, and we choose the characteristic length scale \(\lambda\) and timescale \(\tau\) for this purpose. In general, if these quantities should differ between the reference and target star, then \cref{eq:rhoc} must be modified as
\[
{\delta(\omega \tau)\over \omega \tau} = \int_0^1 K_\rho\ {\delta\log (\rho/\rho_0)}\ \mathrm d x + \int_0^1 K_{c_s}\ {\delta\log (c_s \tau / \lambda)}\ \mathrm d x,\label{eq:rhocunits}
\]
where the characteristic density is specified from the characteristic timescale as \(\sqrt{G \rho_0} = 1/\tau\). For standard inversions, both \(\lambda\) and \(\tau\) must be specified for the true star, and in general may differ from their values in the reference star. \(\lambda\) is chosen to be the stellar radius, so that the integration coordinate is subject to scaling by \(\lambda\). \(\tau\), the dynamical timescale, sets the units of frequency. In our solar model test inversions, the reference and target model are constructed to have the same mass and radius (and thus dynamical timescale \(T_\text{dyn} = \sqrt{R^3 / GM}\)). However, when using the acoustic radial coordinate as the integration variable, it is clear that the acoustic radius is the preferred characteristic timescale, and our solar models do not have the same total acoustic radii. Because of this, we examine inversions with both inputs and outputs scaled in a similar fashion to standard inversions in asteroseismology, where the mass and radius of the true star are not known in advance.

In the modified inversion procedure using free kernels, the acoustic radial coordinate is scaled by \(\tau\) (for which a natural choice would be the acoustic radius \(T\)) rather than \(\lambda\). In \cref{fig:pinversions}a, we show SOLA inversion results obtained from this procedure --- they can be seen indeed to trace the relative Lagrangian differences in the sound speed, rather than Eulerian ones.

For standard inversions of stars other than the Sun, the physical and acoustic radius of the true star is not known a priori. Moreover, for standard inversions, specifying different putative values of \(\tau\) or \(\lambda\) for the target star than it actually possesses induces constant offsets between the inverted values and ground truth \citep{basu_inversions_2003}. However, when the acoustic radial coordinate is used, the inputs, outputs, \emph{and} the integration coordinate of the inversions are all to be scaled by \(\tau\). Because of this, the accuracy of this modified procedure turns out to be insensitive to the precise value of \(\tau\) assumed for the target star. We illustrate this in \cref{fig:pinversions}b. While changes to \(\tau_\text{targ}\) do indeed change the inversion results, these modifications are of exactly the same shape as those induced into the ground-truth structure differences by also rescaling the coordinate axes. Moreover, unlike inversion in the physical radial coordinate, \(\lambda_\text{targ}\) does not need to be specified to set up the inversion problem, as it is not needed to define the integration coordinate for the true star in the first place. Rather, the same value of it (not necessarily the stellar radius) is used for both the reference and the true star, as part of a fixed system of units.

It is instructive to contrast this with how the ground truth and the inversion results for standard Eulerian inversions in the physical radial coordinate respond to differences in \(\lambda_\text{targ}\) and \(\tau_\text{targ}\):

\begin{itemize}
\tightlist
\item
  When varying \(\lambda_\text{targ}\), keeping \(\tau_\text{targ}\), the shape and slope of the ground-truth curve would change (as in \cref{fig:pinversions}b), as each point in the reference model is being compared against a different point in the target model for each candidate value of \(\lambda_\text{targ}\). However, since \(\tau_\text{targ}\) is being held fixed, the left-hand-side of \cref{eq:rhocunits} does not change, and therefore the inversion results (being the same linear combinations of frequency differences) are invariant. As such, the inversion results (which do not change) and ground truth respond differently to different values of \(\lambda_\text{targ}\).
\item
  When varying \(\tau_\text{targ}\), keeping \(\lambda_\text{targ}\) fixed, the shape and slope of the inversion results would change (as in \cref{fig:pinversions}b). This is because, while all the input frequency differences are shifted by the same constant offset upon such a transformation, the inversion coefficients for each localisation kernel do not sum to 1 (instead they are chosen so that each linear combination of kernels has unit integral) and so the inversion result at each target location is modified by a different amount. However, the overall effect on the ground truth structure differences is only a uniform translation in the vertical direction resulting from the change in units (for both quantities). Again, the inversion results and ground truth respond differently to different values of \(\tau_\text{targ}\).
\end{itemize}

As such, \emph{unlike standard Eulerian inversions}, modified p-mode inversions in the acoustic radial coordinate do not require \(\lambda\) and \(\tau\) both to be accurately specified for the true star. The former in fact does not need to be specified at all. As long as the reference model is sufficiently similar to the true star, the latter may be supplied entirely from the reference model, or estimated using e.g.~the large separation \(\Delta\nu\).

\subsection{g-modes: Linearising Post-Main-Sequence Inversions}\label{sec:pms}

\begingroup
\renewenvironment{figure}{\begin{figure*}}{\end{figure*}}

\begin{figure}
\centering
\pandocbounded{\includegraphics[keepaspectratio]{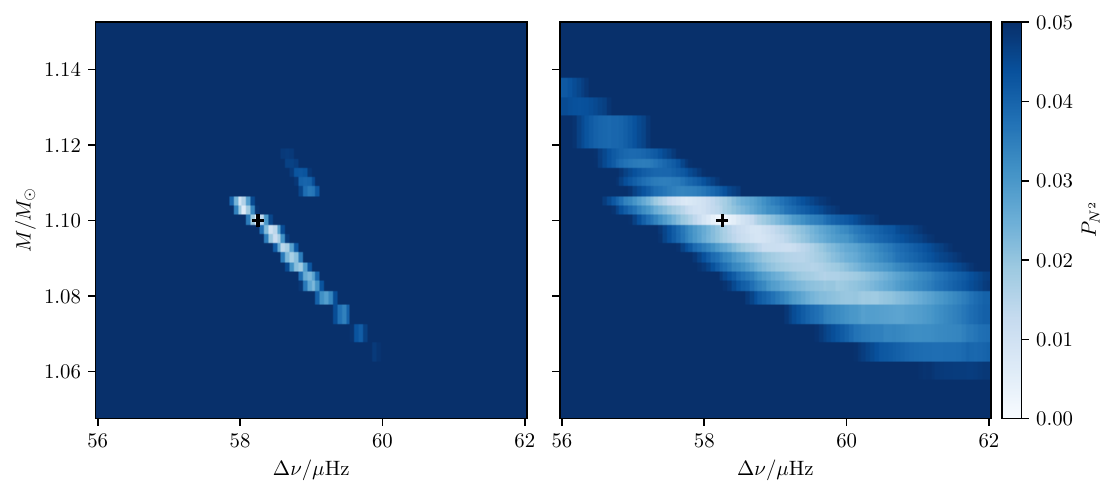}}
\caption{Nonlinearity in subgiant g-mode inversions. A reference subgiant model (whose position is denoted with a plus symbol in both panels) is compared against other stellar models with various masses and large frequency separations \(\Delta\nu\), using the noncommutativity score defined in \cref{eq:nc}. This score is calculated for the Brunt-Väisälä frequency kernels of the \(n_g = 3\) \(\gamma\)-mode, and shown with a colourmap that saturates at a 5\% discrepancy in the shapes of the kernels. In the left panel, we show values evaluated with standard scaling in the fractional radius applied to the Eulerian kernels, while in the right panel we show these computed with modified scaling for modified kernels in the buoyancy radial coordinate. \label{fig:commutativity}}
\end{figure}

\endgroup

Standard asteroseismic inversions break down when applied to g-modes in slowly-pulsating B (SPB) stars \citep{vanlaer_feasibility_2023} and gravitoacoustic mixed modes in sub- and red giants \citep{buchele_linearity_2025}. In both cases, the shapes of the eigenfunctions (and therefore sensitivity kernels) change significantly, both over time evolution, and with respect to other stellar properties (such as mass and composition). These changes are driven both by structural evolution --- e.g.~changes to core sizes and chemical composition profiles --- and, in sub- and red-giants, also by altering the configuration of resonances between p- and g-modes, which couple to yield mixed modes. Such discrepancies between the shapes of the kernels in the reference and target structure induce nonlinear dependences of mode frequencies on structural differences, thereby invalidating the use of \cref{eq:kernel}.

Let us suppose that we have access to the internal structures, sensitivity kernels, and mode frequencies of two similar stellar structures --- which we label models (1) and (2) for the discussion below. If \cref{eq:kernel} should hold, then it should be able to predict both the frequencies of model (1) given the sensitivity kernels of model (2), and the frequencies of model (2) given the senstivity kernels of model (1). If so, then
\[\int \left(K^{(1)}_{i,k} - K^{(2)}_{i,k}\right) \delta u_k (x)\ \mathrm d x\]
would vanish for each mode \(i\) and observable \(k\) --- the two linear operations commute. If this were to hold for arbitrary variations, then the fundamental lemma of the calculus of variations requires that the term in the parentheses should itself vanish everywhere. This being so, we quantify the noncommutativity between the two linear predictions using what we will call the ``noncommutativity score'',
\[P_k = \int \left|K^{(1)}_{i,k} - K^{(2)}_{i,k} \right| \mathrm d x, \label{eq:nc}\]
whose value corresponds to a discrepancy in the predicted fractional mode frequency.

\citet{buchele_linearity_2025} demonstrate that, for subgiant mixed modes, the formal radius of convergence for inversions is primarily determined by the evolution of mixed-mode resonances. In principle, one might circumvent these issues by performing inversions with respect to pure p- and g-modes, e.g.~using isolation schemes like that of \citet{ong_semianalytic_2020}. However, even in the absence of resonances, the shapes of g-mode kernels are highly sensitive both to evolution and to stellar properties. This is because the spatial extent (in terms of fractional radius) of the radiative helium core, and therefore of the g-mode cavity, changes rapidly with stellar evolution, and is highly sensitive to the stellar mass. In the left panel of \cref{fig:commutativity}, we can see that \cref{eq:kernel} remains commutative (with a small noncommutativity score), and therefore that standard linear inversions remain feasible, only in a small region of parameter space around a reference model (shown with a plus symbol), corresponding to where the radiative core occupies the same fractional volume of the target model as the reference model.

Suppose instead we were capable of measuring the period spacing of the true star, and the frequencies of the g-modes underlying its mixed modes. We would then be able to supply the characteristic timescale from the former, and employ modified kernels in the buoyancy radial coordinate for inversions using the latter as inputs. Doing so would also render inversions linear over a broader range of parameter space, as we see in the right panel of \cref{fig:commutativity}, even where the radiative cores of the reference and target models are of different sizes.

\section{Summary and Future Prospects}\label{summary-and-future-prospects}

Rather than merely constraining forward models of stellar interiors, asteroseismic inversions permit us to measure their structure directly. They were a key tool in allowing us to understand the solar interior, and their application to other kinds of stars would be highly valuable for our efforts to understand them as well.

However, it has become increasingly evident that existing techniques for asteroseismic inversions must be significantly modified for use on stars that are significantly different from the Sun --- and most stars in which we are able to measure normal-mode frequencies are not Sun-like. Performing these inversions in coordinates other than the physical radius has long been thought to be the most obvious such modification. Yet, until this work, this has only been possible for studying stellar structure when using the mass coordinate, as in \citet{jcd_solar_1997}, rather than in coordinates that match the asymptotic behaviour of the normal modes \citep{tassoul_asymptotic_1980, tassoul_second_1990, roxburgh_asymptotic_1994}. Even more recent asymptotic inversions \citep[as in e.g.][]{briganti_predictions_2025, guo_glitch_2025, guo_inferring_2025} rely on asymptotic approximations to what we have shown to be sensitivity kernels in the Eulerian structure differences, rather than numerical kernels correctly describing structure differences taken at matching structural coordinates.

We have now developed the mathematical and methodological infrastructure necessary for generally describing such numerical kernels. As such, we have answered the question posed by the title of this work in the affirmative: we \emph{can}, indeed, now prosecute asteroseismic structure inversions using coordinate systems that are themselves structure-dependent. Along the way, we have also uncovered multiple reasons why we \emph{should} do so:

\begin{itemize}
\tightlist
\item
  Although techniques adapted from helioseismology have been successfully applied to Sun-like p-mode pulsators, we have shown that p-mode inversions in the acoustic radial coordinate would still be advantagous over those in the physical radius. They eliminate the methodological requirement that both the mass and radius of the true star be first measured in advance. While some value for the acoustic radius of the true star does need to be assumed, the inversion results are insensitive to the exact value used. Moreover, the radius of the true star does not even need to be constrained in advance --- the only length scale that needs to be specified is that of the overall system of units. Finally, the modified kernels have different localisation properties (particularly close to the centre and the surface) than standard Eulerian kernels. Sensitivity to the centre of the star in fact permits these modified kernels to infer from inversions that the target structure possesses a different total radius from the reference model, which by construction cannot be achieved by Eulerian inversions in the fractional radius.
\item
  Even our best-fitting models of g-mode pulsators will, in all likelihood, have differently-sized and differently massive convective or radiative cores than the true stars actually possess. Indeed, studying the physical processes that determine the sizes of these cores is one main reason why we might want to conduct asteroseismic inversions in the first place. We have shown that our modified kernels directly address the \emph{reasons} why linear structure inversions break down upon application to to g-modes. Therefore, obtaining such measurements by evaluating structural differences at matching fractional buoyancy radii, rather than physical radii or mass, may well be the only meaningful path to pursuing linear structure inversions using g-modes.
\end{itemize}

These two classes of modes suffice to cover nearly the entire asteroseismic Hertzsprung-Russell diagram --- for which, we submit, data availability rather than methodological obstructions may now be the limiting factor in whether or not asteroseismic inversions can be conducted.

One exception to this is the case of modes of mixed character, such as seen in subgiants and red giants. There, nonlinear effects arise when the contributions from the off-diagonal matrix elements render \cref{eq:rsperturb}, and therefore \cref{eq:kernel}, no longer directly usable \citep{buchele_linearity_2025}. Linear inversions, separately in the acoustic and buoyant radii, remain possible in the isolated basis set of pure p- and g-modes \citep[per][]{ong_semianalytic_2020, ong_rotation_2022}, at the cost of discarding measurements of the strength of the coupling between these p- and g-modes. In future work, we plan to incorporate additional information about stellar structure, that may be encoded in these coupling strengths, into the inversion procedure.

\begin{acknowledgements}

We thank Conny Aerts, Jørgen Christensen-Dalsgaard, Margarida Cunha, Sebastian Deheuvels, and Saskia Hekker for productive discussions about prospects for mixed- and g-mode inversions. We also thank the anonymous referee for their careful review. J.M.J.O. acknowledges support from NASA through the NASA Hubble Fellowship grant HST-HF2-51517.001, awarded by STScI, and from the Australian Research Council (FL220100117 and FT200100871). STScI is operated by the Association of Universities for Research in Astronomy, Incorporated, under NASA contract NAS5-26555. W.B.H acknowledges support from the National Science Foundation Graduate Research Fellowship Program under Grant Nos. 1842402 and 2236415. V.V. gratefully acknowledges support from the Research Foundation Flanders (FWO) under grant agreement N°1156923N (PhD Fellowship).

Our \texttt{jax}-accelerated implementation of the numerical routines described in this work are made freely available on Zenodo at \dataset[10.5281/zenodo.16537658]{\doi{10.5281/zenodo.16537658}}.

\software{NumPy \citep{numpy}, SciPy stack \citep{scipy}, AstroPy \citep{astropy:2013, astropy:2018}, Pandas \citep{pandas}, \mesa~\citep{mesa_paper_1, mesa_paper_2, mesa_paper_3, mesa_paper_4, mesa_paper_5}, \gyre~\citep{townsend_gyre_2013}, \texttt{jax} \citep{jax}.}

\end{acknowledgements}

\appendix

\section{Coordinate Expressions for the Volterra operator}\label{coordinate-expressions-for-the-volterra-operator}

\subsection{Convergence of Neumann Series}\label{sec:convergence}

The convergence of iterative procedures like that of \cref{eq:iteration} is not guaranteed. To see why, consider how such a solution might be found numerically: we might discretise the integral for the total action as
\[
\delta S \approx \sum_k \sum_a K_k(x_a) \delta f_k \Delta x_a \equiv \mathbf{K}^T \mathbf{M}_x\mathbf{d},\label{eq:matrix}
\]
where \(\mathbf{K}^T\) is a single row vector containing concatenated values of the kernels, \(\mathbf{M}_x\) is a diagonal matrix whose entries are the approximate integral measure at each discrete mesh point \(x_a\) indexed by \(a\), and \(\mathbf{d}\) is a single column vector containing concatenated values of the perturbations \(\delta f\). The integral in \cref{eq:lagrange-to-euler} is replaced by a sum
\[
\delta \tilde f_k(y(x_a)) = \sum_{l,b} \left(\delta_{kl}\delta_{ab} - {\mathrm d\tilde f_k \over \mathrm d y}(x_a){\partial y(x_a) \over \partial f_l(x_b)}\right) \delta f_l(x_b),
\]
which can be compactly written as a matrix product \(\tilde{\mathbf{d}} = \left(\mathbb{I} - \mathbf{T}\right)\mathbf{d}\), where \(\mathbb{I}\) is the identity matrix. This yields
\[
\tilde{\mathbf{K}} = \mathbf{M}_y^{-1}\left(\mathbb{I} - \mathbf{T}\right)^{-T}\mathbf{M}_x\mathbf{K}.\label{eq:matinv}
\]
Solutions by Neumann series correspond to series expansions in powers of \(\mathbf{T}^T\), which converge only if the norm of \(\mathbf{T}^T\) is small.

In practice, however, such iterative schemes tend to converge fairly rapidly. Let us use the same definition as in \autoref{sec:changes} of \(y(x)\) being of the form \(h(x) / H\), where \(h = \int_0^x J\ \mathrm d x'\). Under a fixed outer boundary, \(H = h(1)\),
\[
{\delta y(x) \over \delta f(x')} = {1 \over H}{\delta H \over \delta f(x')}\left(\theta(x - x') - y(x)\right), \label{eq:funcdery}
\]
where \(\theta(a)\) is the Heaviside step function (which returns 1 when \(a > 0\), \(1\over2\) when \(a = 0\), and 0 otherwise). Alternatively, if we were to assume a free boundary, then the second term in the parentheses is dropped.

Applying the identity \cref{eq:funcdery} to the acoustic radius \(y(x) = t(x)/T\) gives
\[
{\delta y(x) \over \delta \log c_s(x')} = -{R \over T c_s(x')}\left[\theta(x - x') - y(x)\right],\label{eq:funcdert}
\]
with the second term suppressed under a free boundary. We may discretise this, for use with \cref{eq:matinv}, to yield
\[
{\partial y_i \over \partial \log {c_s}_j} = - {R \over T {c_s}_j} \left[\theta(x_i - x_j) - y_i\right]\Delta x_j = \left[\theta(x_i - x_j) - y_i\right]\Delta y_j.
\]
Taking the \(\ell_1\) operator norm of \(\mathbf{T}\) then gives
\[
|\mathbf{T}| \lesssim \max_y \left|y {\mathrm d \log c_s(y)\over \mathrm d y}\right|
\]
with a free boundary, and \(\sim \max \left|{\mathrm d \log c_s\over \mathrm d y}\right|\) for a fixed boundary. Numerical inspection of stellar models indicates this to be \(\ll 1\) for prototypical p-mode pulsators, such as Sun-like main-sequence and red-giant stellar models. Similarly, we find the analogous g-mode matrix norm of \(\max \left|y {\mathrm d \log N^2\over \mathrm d y}\right|\) also to be small for the g-mode-pulsator stellar models examined in \citet{vanlaer_feasibility_2023}.

\subsection{Expressions for modified kernels}\label{expressions-for-modified-kernels}

These explicit expressions for the matrix elements of the Volterra operator permit us in turn to solve for the modified kernels in a few special cases. We will focus on specifically \(\tilde K_{c_s,\rho}(t/T)\) and \(\tilde K_{N^2,\rho}(b/B)\) under free outer boundaries, since these are of particular physical interest and have featured heavily in the main text of the paper.

For \(\tilde K_{c_s,\rho}\) in the acoustic radial coordinate, inserting \cref{eq:funcdert} gives, with a free outer boundary, that
\[
\begin{aligned}\int 
& K_{c_s,\rho}(x) \delta \log c_s(x) \mathrm d x + (\text{terms for $\delta\rho$})\\
=& \int_0^1 \tilde K_{c_s,\rho}(y)\left[ \delta \log c_s(x) + {\mathrm d \log c_s \over \mathrm d y} \int_0^x {R \over T c_s(x')}\delta \log c_s(x')\mathrm d x' \right]\mathrm d y. \\
&+ \int_0^1 \tilde K_{\rho, c_s}(y) {\mathrm d \log \rho \over \mathrm d y} \int_0^x {R \over T c_s(x')}\delta \log c_s(x')\ \mathrm d x'\mathrm dy \\
&+ (\text{terms for $\delta\rho$})
\end{aligned}
\]
Exchanging the order of integration on the right-hand-side, observing that \(\tilde K_{\rho,c_s} = {\mathrm d x \over \mathrm d y} K_{\rho,c_s}\), and applying the fundamental lemma of the calculus of variations yields
\[
\begin{aligned}
{\mathrm d y \over \mathrm d x}& \tilde K_{c_s,\rho}(y(x)) + {1 \over c_s(x)} \int_x^1 {\mathrm d y \over \mathrm d x'} \tilde K_{c_s,\rho}(y(x')) {\mathrm d c_s(x') \over \mathrm d x'} \mathrm d x' \\
&= K_{c_s,\rho}(x) - {1 \over c_s(x)} \int_x^1 K_{\rho, c_s}(x') {c_s\over\rho} {\mathrm d\rho \over \mathrm d x'}\ \mathrm dx' \\
&\equiv J_{c_s,\rho}(x).\label{eq:appcs1}
\end{aligned}
\]
We can cast this integral equation in \(\tilde K_{c_s,\rho}\) as an equivalent differential equation solvable using standard techniques, and we note that this expression also implies an outer boundary condition that \(J_{c_s,\rho}(x = 1) = K_{c_s,\rho}(x = 1) = \left.{\mathrm d y \over \mathrm d x} \tilde K_{c_s,\rho}(y(x))\right|_{x=1}\). Putting these together, we obtain at last that
\[
\tilde K_{c_s,\rho}(y(x)) = {\mathrm d x \over \mathrm d y} \left(J_{c_s,\rho}(x) - \int_x^1 J_{c_s,\rho}(x') {\mathrm d \log c_s \over \mathrm d x'} \mathrm d x'\right),\label{eq:appcs2}
\]
which we further simplify in \autoref{sec:centre}. A similar calculation for the Brunt-Väisälä frequency kernels and the buoyancy radius yields that
\[
\tilde K_{N^2,\rho}(y(x)) = {\mathrm d x \over \mathrm d y} \left(J_{N^2,\rho}(x) + {1 \over x}\int_x^1 x' J_{N^2,\rho}(x') {1\over2}{\mathrm d \log N^2 \over \mathrm d x'} \mathrm d x'\right),\label{eq:Ktilde-buoyancy}
\]
with
\[
J_{N^2,\rho} = K_{N^2,\rho} + {N \over 2x} \int_x^1 K_{\rho, N^2} {x' \over N} {\mathrm d \log \rho \over \mathrm d x'} \mathrm d x',
\]
and so (similarly to \cref{eq:Iterms}) that
\[
\begin{aligned}
\tilde K_{N^2,\rho}(y(x)) = &{\mathrm d x \over \mathrm d y} \left(K_{N^2,\rho} + {1 \over 2x} \int_x^1 K_{\rho, N^2}  {\mathrm d \log \rho \over \mathrm d \log x'} \mathrm d x'\right.\\
&+\left.{1 \over 2x}\int_x^1 K_{N^2,\rho} {\mathrm d \log N^2 \over \mathrm d \log x'} \mathrm d x'\right).
\end{aligned}
\]

\section{Buoyancy-Frequency Kernels}\label{sec:N2kernels}

\citet{vanlaer_feasibility_2023} describe the construction of sensitivity kernels using the Brunt-Väisälä frequency as one of the dynamical variables. However, they noted that their implementation of these kernels was prone to numerical errors, particularly in the \(N^2\)-\(\rho\) kernel pair, and speculated that this originates from numerical instabilities in the differentiation scheme they used to construct these kernels.

\begin{figure}
\centering
\pandocbounded{\includegraphics[keepaspectratio,alt={Density kernels at constant N\^{}2, K\_\{\textbackslash rho, N\^{}2\}, computed either using finite-difference derivatives as in @vanlaer\_feasibility\_2023 (dashed curve), or using linear combinations of other eigenfunctions, as in  (solid curve), for computing \textbackslash mathrm d \textbackslash xi\_r / \textbackslash mathrm d r and higher derivatives. Avoiding the use of repeated finite-difference derivatives almost completely suppresses numerical instabilities --- compare their Fig. A1 --- rendering the N\^{}2-\textbackslash rho kernel pair numerically viable for use in this work.}]{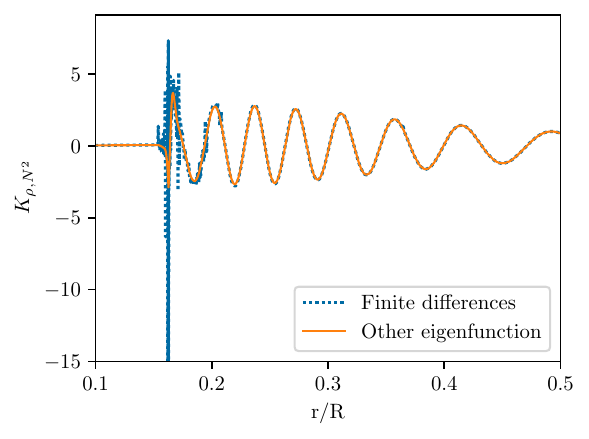}}
\caption{Density kernels at constant \(N^2\), \(K_{\rho, N^2}\), computed either using finite-difference derivatives as in \citet{vanlaer_feasibility_2023} (dashed curve), or using linear combinations of other eigenfunctions, as in \cref{eq:deriv} (solid curve), for computing \(\mathrm d \xi_r / \mathrm d r\) and higher derivatives. Avoiding the use of repeated finite-difference derivatives almost completely suppresses numerical instabilities --- compare their Fig. A1 --- rendering the \(N^2\)-\(\rho\) kernel pair numerically viable for use in this work.\label{fig:finitedifferences}}
\end{figure}

In the specific case of the transformation from \(c^2\)-\(\rho\) to \(N^2\)-\(\rho\) kernels, we find that finite-difference derivatives were repeatedly taken of the radial Lagrangian displacement eigenfunction \(\xi_r\) when effecting this tranformation numerically in \citet{vanlaer_feasibility_2023}. However, each application of finite-difference numerical differentiation introduces both roundoff and truncation error, which is amplified by successive differentiations --- this is the source of the numerical instability noted in that work. Such instabilities can be avoided by exploiting the internal structure of the pulsation equations, which relates derivatives of each eigenfunction to linear combinations of each of the rest. In particular, the identity
\[{\mathrm d \over \mathrm d r}\xi_r = \left({g \over c_s^2} - {2 \over r}\right) \xi_r - \alpha_\gamma {P' \over \rho c_s^2} + {\ell(\ell + 1) \over r} \xi_h\label{eq:deriv}\]
may be applied repeatedly in lieu of finite-difference derivatives --- note that an additional factor of \(\alpha_\gamma\) (\(= 0\) for \(\gamma\)-modes, but \(= 1\) for standard normal modes) also has to be introduced for numerical consistency when applying it to differentiate \(\gamma\)-mode eigenfunctions. We show in \cref{fig:finitedifferences} the numerical kernels that result from using this implicit differentiation scheme. It is clear that the numerical instabilities are almost entirely eliminated. With them rectified, we find the differences between forward-modelled frequency differences computed using \(c^2\)-\(\rho\) kernels vs.~\(N^2\)-\(\rho\) kernels, as originally reported in \citet{vanlaer_feasibility_2023}, also to be eliminated.

We were able to achieve similar improvements in numerical stability for obtaining \(N^2\)-\(c_s^2\) from \(c^2\)-\(\rho\) kernels: rather than explicitly solving the integro-differential equation that is required to effect the transformation, we instead were able to formulate the problem as a higher-order integral equation, analogously to that described in the main text, from which we produced iterative solutions. Specifically, defining \(Y = g K_{N^2, c_s^2}\), we have
\[Y(r) = \int_{r}^R \left(4 \pi G \rho r'^2 \int_{r'}^R Y(r'') g^{-1} r''^{-2} \left({N^2 \over g} - {g \over c_s^2}\right)\mathrm d r'' - K_{\rho, c_s^2} \right)\mathrm d r'.\]
We found that using \(K_{N^2, \rho}\) as an initial guess for constructing self-consistent iterative solutions by Neumann series, analogously to \cref{eq:iteration}, yielded convergence within a very small number of iterations.

\section{Buoyancy-radius kernels in convection zones}\label{sec:deltafuncts}

\begin{figure}
\centering
\pandocbounded{\includegraphics[keepaspectratio,alt={g-mode kernels for the same mode and model as shown in the right half of , plotted in the physical radial coordinate. Line colours and stroke widths have the same meaning as in . The location of convection zones, including the convective core, are indicated with the blue shaded region. }]{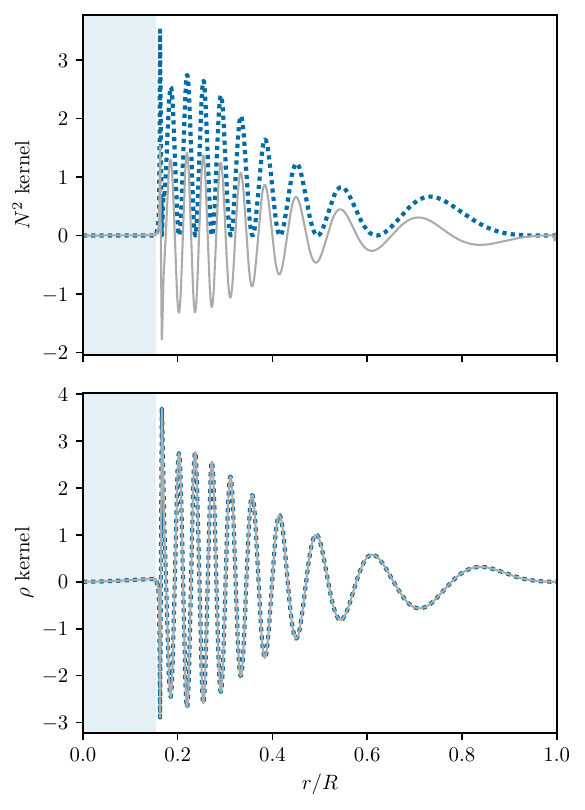}}
\caption{g-mode kernels for the same mode and model as shown in the right half of \cref{fig:rescaled}, plotted in the physical radial coordinate. Line colours and stroke widths have the same meaning as in \cref{fig:rescaled}. The location of convection zones, including the convective core, are indicated with the blue shaded region. \label{fig:gmodecz}}
\end{figure}

In principle, within each convection zone, the Brunt-Väisälä frequency vanishes, and it therefore takes up no width in the buoyancy radial coordinate. Accordingly, when changing integration variables from the physical to the buoyant radial coordinate, \cref{eq:kernel} needs to be modified as
\[
{\delta\omega\over\omega} \sim \sum_j \left(\int_{N^2 \ne 0} \tilde K_j(b) \delta \tilde u_j(b)\ \mathrm d b + \sum_k \int_{x_{i,k}}^{x_{o,k}} \tilde K_j(x) \delta \tilde u_j(x)\ \mathrm d x\right),\label{eq:cz}
\]
where the sum indexed by \(j\) runs over physical quantities, the integral over the buoyancy coordinate is understood to avoid the convection zones, and the sum over \(k\) runs over convection zones.

While this piecewise procedure described is necessary if \(N^2 = 0\) exactly inside a convection zone, numerical stellar models (e.g.~those produced by \mesa) possess convection zones in which \(N^2\) is nonzero, but of very low (physically negligible) absolute value. Nonetheless, for our numerical purposes, this allows us to define a modified buoyancy coordinate as in \cref{eq:bmod}. As described in the main text, this modified coordinate preserves the structure of \cref{eq:b} in radiative zones, but remains monotonic throughout the star even in convection zones, allowing us to compute \(\tilde{K}\) in a well-defined manner.

To illustrate the convection-zone contributions from \cref{eq:cz}, we show \(K(x)\) in \cref{fig:gmodecz}, highlighting all convection zones in the stellar model (both from the convective core, and from near-surface convective shells) in blue shaded regions. In these parts of the star, all kernels can be seen to be close to zero, so that contributions from the convection-zone terms in \cref{eq:cz} can be seen to be much less significant than from inside the radiation zone. We also show \(\tilde K(x)\) computed in this fashion with the grey curve --- as with the Eulerian kernels, the contribution from the convection zones is also very small.

This property follows from asymptotic theory \citep[e.g.][]{tassoul_asymptotic_1980, tassoul_second_1990}, as the local amplitudes of g-mode eigenfunctions are determined by the Brunt-Väisäla frequency, and therefore should vanish where it vanishes. Since the structure kernels are computed as bilinear operators acting on these eigenfunctions, they too should vanish there in the asymptotic regime.

\begin{figure}
\centering
\pandocbounded{\includegraphics[keepaspectratio,alt={n\_g=3 g-mode kernels for fiducal subgiant model considered in , plotted in the same manner as . }]{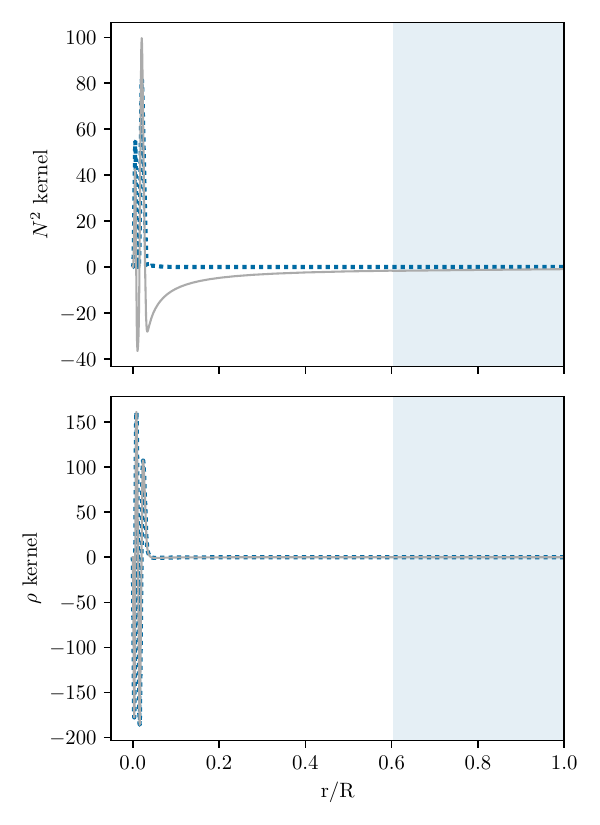}}
\caption{\(n_g=3\) g-mode kernels for fiducal subgiant model considered in \autoref{sec:pms}, plotted in the same manner as \cref{fig:gmodecz}. \label{fig:gmodecz2}}
\end{figure}

However, these contributions are small even for modes that are not necessarily in the asymptotic regime. For example, we show the \(n_g=3\) kernels of our reference subgiant model (from \autoref{sec:pms}) in \cref{fig:gmodecz2}. Despite this mode not being of particularly high radial order, its kernels still take values much closer to 0 in the convection zones than in the radiative interior. In general, we find that this appears to be a good approximation so long as the shape of the mode eigenfunction is not predominantly affected by mode trapping in interfacial layers \citep[of the kind described in][]{pontin_wave_2020}. This being the case, we neglect these convection-zone terms in our analysis presented in the main paper.

\section{Central sensitivity of acoustic-radius kernels}\label{sec:centre}

Unlike the Eulerian kernels, our modified kernels in the acoustic radial coordinate exhibit sensitivity to the centre of the star. We examine this in more detail in this appendix. In particular, it is possible to rearrange \cref{eq:appcs1,eq:appcs2} so as to permit the integrals appearing there to run from 0 to \(x\), instead of from \(x\) to 1. Doing so, and interchanging the order of integration for a double integral, yields at last that

\begin{widetext}
\begin{equation}
\begin{aligned}
\tilde K_{c_s,\rho}(y(x)) &= {\mathrm dx\over \mathrm dy} \left(\underbrace{K_{c_s,\rho}(x) + {1 \over c_s} \left[\int_0^x K_{\rho, c_s}(x') {c_s \over \rho} {\mathrm d \rho\over \mathrm d x'}\ \mathrm d x' - I_1\right]}_{=J_{c_s,\rho}(x')} + \int_0^x J_{c_s,\rho}(x') {\mathrm d \log c_s \over \mathrm d x'} \mathrm d x' - I_2\right)\\
&= {\mathrm dx\over \mathrm dy} \left(K_{c_s,\rho}(x) + {1 \over c_s} \int_0^x K_{\rho, c_s}(x') {c_s \over \rho} {\mathrm d \rho\over \mathrm d x'}\ \mathrm d x' + \int_0^x K_{c_s,\rho}(x') {\mathrm d \log c_s \over \mathrm d x'} \mathrm d x' + \int_0^x \left[\int_{x'}^x{1 \over c_s} {\mathrm d \log c_s \over \mathrm d x''} \mathrm d x''\right] K_{\rho, c_s}(x') {c_s \over \rho} {\mathrm d \rho\over \mathrm d x'}\ \mathrm d x'  - {1\over c_s} I_1 -  I_2\right)\\
&= {\mathrm dx\over \mathrm dy} \left(K_{c_s,\rho}(x) + \int_0^x K_{\rho, c_s}(x') {\mathrm d \log \rho\over \mathrm d x'}\ \mathrm d x' + \int_0^x K_{c_s,\rho}(x') {\mathrm d \log c_s \over \mathrm d x'} \mathrm d x'  - {1\over c_s} I_1 -  I_2\right)\\
& \equiv \overline K - {\mathrm d x \over \mathrm d y}\left({1\over c_s} I_1 -  I_2\right),\label{eq:Iterms}
\end{aligned}
\end{equation}
\end{widetext}

where \(I_1 = \int_0^1 K_{\rho, c_s}(x') {c_s \over \rho} {\mathrm d \rho\over \mathrm d x'}\ \mathrm d x'\) and \(I_2 = \int_0^1 J_{c_s,\rho}(x') {\mathrm d \log c_s \over \mathrm d x'}\ \mathrm d x'\) are constants. By inspection, as written in this form, only the terms proportional to \(I_1\) and \(I_2\) remain nonzero as \(x\to0\), and so clearly it is these terms which are responsible for central sensitivity of the modified kernels. Note that \({\mathrm d x \over \mathrm d y} = {\tau \over \lambda}c_s\); in other words, \(\tilde K_{c_s,\rho}\) for each mode can be written as a linear combination of some \(\overline K\) with shape specific to each mode, an overall constant contribution (proportional to \(I_1\)), and a contribution proportional to \(c_s\) (also proportional to \(I_2\)). Correspondingly,
\[
\begin{aligned}
{\delta (\omega \tau) \over \omega \tau} \sim & \int \overline K \delta \log\left(c_s \tau / \lambda\right) \mathrm d y + \int \tilde K_\rho \delta \log\left(\rho / \rho_0\right) \mathrm d y \\ &+ \underbrace{I_1 \int \delta \log\left(c_s \tau / \lambda\right) \mathrm d y}_{= I_1 \left<\delta \log c_s\right>} + \underbrace{I_2 \int {\tau \over \lambda} c_s \delta \log\left(c_s \tau / \lambda\right) \mathrm d y}_{= I_2 \delta (R/\lambda)}.
\end{aligned}
\]
That is to say, the fact that kernels \(\tilde K_{c_s,\rho}\) are sensitive to the centre is what allows the sound-speed profile returned from the inversion procedure to be relatively larger or smaller on average than the reference model (\(I_1\)), and/or to imply a different total stellar radius (\(I_2\)).

\begin{figure}
\centering
\pandocbounded{\includegraphics[keepaspectratio,alt={Free-boundary modified kernels without (blue dotted curve) and with (orange solid curve) the constraint that the nondimensional stellar radius is unchanged by the sound-speed perturbation.}]{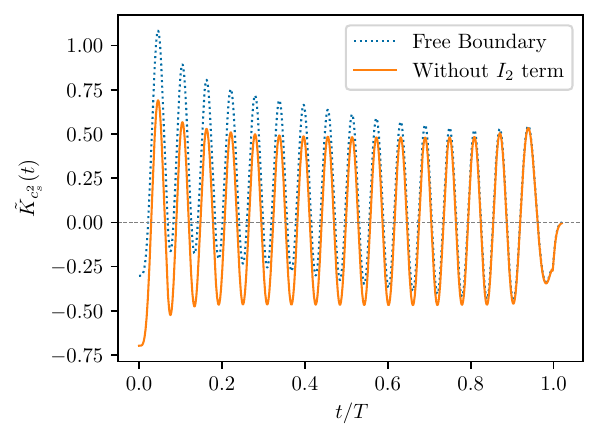}}
\caption{Free-boundary modified kernels without (blue dotted curve) and with (orange solid curve) the constraint that the nondimensional stellar radius is unchanged by the sound-speed perturbation.\label{fig:noI2}}
\end{figure}

The second term, proportional to \(I_2\), can be suppressed by setting \(\lambda = R\) separately for both the target and reference structure, as is standard practice in Eulerian inversions --- this amounts to a constraint that, with the target and reference systems separately nondimensionalised, the nondimensional stellar radius is unchanged by the sound-speed perturbation. This is a similar situation to how a term proportional to \(4\pi r^2 \rho\) is often separated from the conventional density kernel (as a so-called ``complementary function'') in Eulerian inversions, thereby changing its shape, in order to impose a constraint that the nondimensional total mass is unchanged by the density perturbation. Doing so in this case changes the shape of the remaining terms for \(\tilde K\), yielding a kernel whose shape is, strikingly, apparently centered on 0 (\cref{fig:noI2}). This property may potentially improve its localisation capabilities, at the expense of requiring \(R\) to be determined for the target star rather than being fixed by the system of units, as our discussion in the main text otherwise assumes.

\bibliography{biblio.bib, custom.bib}

\end{document}